\documentclass[12pt]{article}

\usepackage{graphicx}

\usepackage{times}

\newenvironment{sciabstract}{%
\begin{quote} \bf}
{\end{quote}}

\title{Electrostatic forces above graphene nanoribbons and edges interpreted as partly hydrogen-free}

\author
{Sebastian Schneider,$^{1}$ Regina Hoffmann-Vogel$^{2,3\ast}$\\
\\
\normalsize{$^{1}$Physikalisches Institut, Karlsruhe Institute of Technology, Wolfgang-Gaede-Str. 1,
}\\
\normalsize{76128 Karlsruhe, Germany.}\\
\normalsize{$^{2}$Department of Physics, University of Konstanz, Universit\"atsstrasse 10, 78464 Konstanz,}\\
\normalsize{Konstanz, Germany.}\\
\normalsize{$^{3}$Institut f\"ur Physik und Astronomie, Universit\"at Potsdam, Karl-}\\
\normalsize{Liebknecht-Str. 24-25, 14476 Potsdam, Germany.}\\
\\
\normalsize{$^\ast$Correspondence to: hoffmannvogel@uni-potsdam.de.}
}

\date{}

\begin{document} 


\baselineskip24pt


\maketitle


\begin{sciabstract}
  Graphene nanoribbons' electronic transport properties strongly depend on the type of edge, armchair, zigzag or other, and on edge functionalization that can be used for band-gap engineering. For only partly hydrogenated edges interesting magnetic properties are predicted. Electric charge accumulates at edges and corners. Scanning force microscopy has so far shown the centre of graphene nanoribbons with atomic resolution using a quartz crystal tuning fork sensor of high stiffness. Weak long-range electrostatic forces related to the charge accumulation on the edges of graphene nanoribbons could not be imaged so far. Here, we show the electrostatic forces at the corners and edges of graphene nanoribbons are amenable to measurement. We use soft cantilevers and a bimodal imaging technique to combine enhanced sensitivity to weak long-range electrostatic forces with the high resolution of the second-frequency shift. Additionally, in our work the edges of the nanoribbons are mainly hydrogen-free, opening to the route to investigations of partly hydrogenated magnetic nanoribbons.
\end{sciabstract}


\section{Introduction}

Graphene nanoribbons are promising candidates for future electronic devices \cite{chen07p1,han07p1}. By influencing the ribbon width on the order of a few nm, the band gap can be engineered \cite{chen07p1,han07p1,wagner13p1}. In addition, the edges of graphene nanoribbons strongly influence the ribbon's electronic transport properties \cite{li08p1,wakabayashi10p1,liu16p1,maksimov13p1,orlof13p1}. Indeed, the current flow in graphene nanoribbons is confined to the edges \cite{allen16p1}. Electric charge accumulates at edges and corners and charge enhancement factors of around 10 are predicted \cite{wang10p1}. The electrostatic state of the edges is important for the electronic properties of graphene nanoribbons, because it induces a doubling of the edge states \cite{silvestrov08p1}. In addition the magnetic properties of graphene nanoribbons sensitively depend on doping \cite{gao18p1}.

Scanning force microscopy dynamic measurement modes can determine the internal atomic structure of a molecule \cite{gross09p1,moreno14p1,dienel15p1,hoffmann-vogel18p1}. Interconnected graphene nanoribbons \cite{dienel15p1}, graphene nanoribbons with internal doping and edge functionalization \cite{kawai18p1,kawai15p1} have been investigated using tuning fork sensors. Tuning fork sensors have a much larger stiffness compared to cantilevers, i.e. they are generally much less sensitive to forces and allow for a closer approach to the molecule while avoiding a snap-to-contact and allow a high spatial resolution \cite{giessibl03p1}. In turn, cantilevers are generally more sensitive to weak long-range forces such as electrostatic forces \cite{lantz03p1,perez16p1,spaeth17p1}. Bimodal oscillation constitutes an extension to dynamic SFM methods, where the cantilever is additionally oscillated at its second harmonic resonance frequency \cite{li09p1,kawai09p1,kawai10p1}. The shift of the second resonance is more sensitive to short-range forces, leading to an increased spatial resolution also for cantilevers \cite{kawai09p1}.

\begin{figure}[h]
\centerline{
\includegraphics[height=3cm]{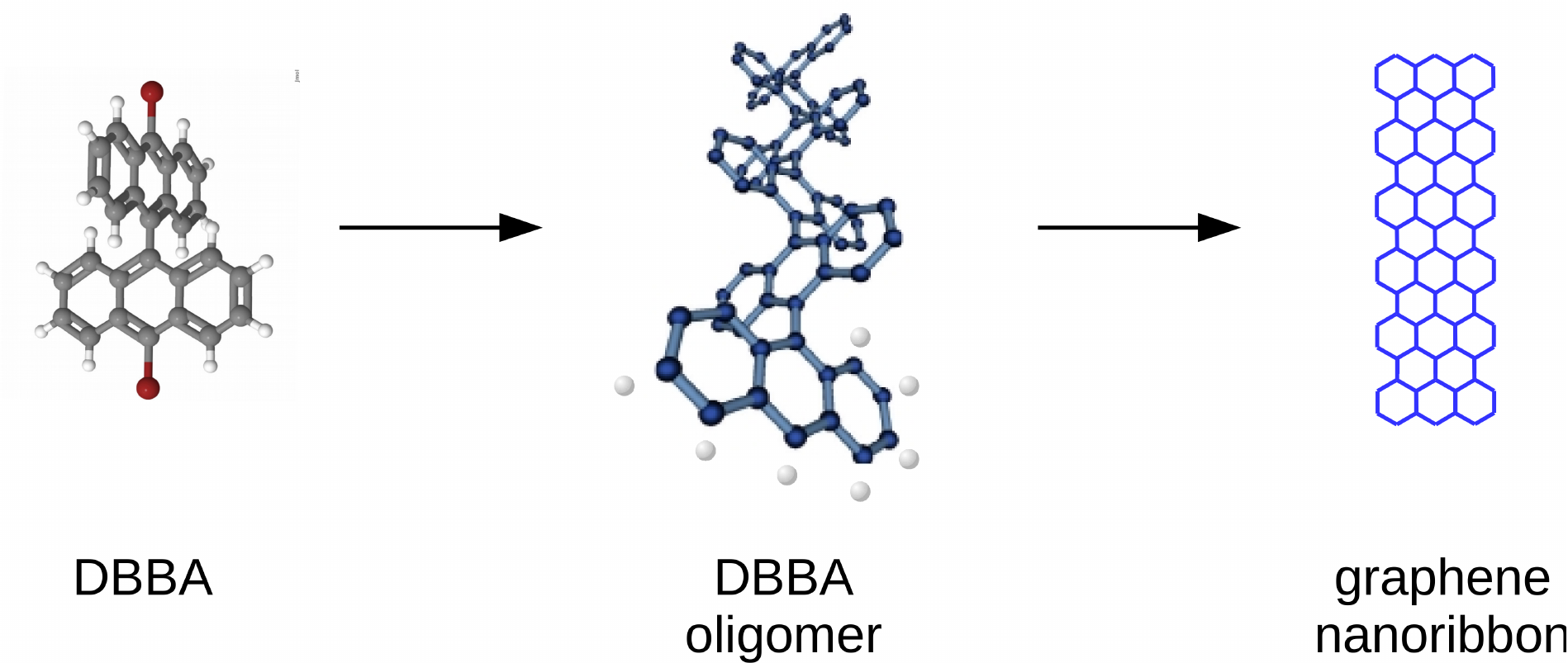}}
    \label{fig0}
\end{figure}

\paragraph*{Fig. 1: Overview of the formation process of graphene nanoribbons via on-surface polymerization.} First DBBA precursor molecules are used (grey: carbon, red: bromine, white: hydrogen). These form oligomers during the on-surface process. In these the carbon atoms at the edges are protected by hydrogen atoms (white spheres). For the sake of clarity of the scheme, only seven hydrogen atoms have been drawn. By further heating, hydrogen atoms are removed, chemical bonds between internal carbon atoms are established and graphene nanoribbons form.

\newpage

Graphene nanoribbons can be prepared on a catalytic metal surface using on-surface synthesis based on Ullmann coupling starting from the precursor molecule 10,10'-dibromo-9,9'-bianthryl (DBBA) \cite{cai10p1}. The overall structure of the resulting graphene nanoribbons is known from literature, from the chemistry of Ullmann coupling and the subsequent Scholl reaction. During the formation of graphene nanoribbons starting from DBBA by heating in vacuum on a Au(111) surface, Br is removed from the molecule and the molecules aggregate and form polymers called DBBA oligomers, see Figure \ref{fig0}. In the DBBA oligomers there are still hydrogen atoms present protecting the oligomers from forming graphene nanoribbons. Further heating removes these hydrogen atoms. It is generally believed that hydrogen is present on the graphene nanoribbons edges at the time of imaging (see \cite{ruffieux16p1} Supplementary Information, and \cite{he13p1}). The presence or absence of hydrogen at the edges is crucial for magnetism in graphene nanoribbons \cite{mishra20p1}. Only for partly hydrogenated graphene nanoribbons magnetism is expected. So far the only route to hydrogen-free graphene nanoribbons is by atomic manipulation, a slow and strongly localized process.

Here, we show bimodal images of graphene nanoribbons with a graphene nanoribbon attached to the tip. We measure with high sensitivity and model the electrostatic forces above the graphene nanoribbons. Additionally, we observe defective sites on the graphene nanoribbon edges. Our measurements show that for our sample preparation and measurement conditions the graphene nanoribbons are largely hydrogen-free. We attribute the difference to literature results to our larger temperature of observation and the use a flow cryostate instead of a bath cryostate. Having hydrogen-free graphene nanoribbons at hand allows to study partial hydrogenation by dosing with additional hydrogen.

\section{Experimental section}
\subsection{On-surface preparation of graphene nanoribbons}

We prepared graphene nanoribbons on the Au(111) surface by on-surface synthesis following the method described in \cite{cai10p1}: As substrates we use Au(111) on mica (Phasis, Plan-les-Oates, Switzerland). We have repeatedly re-cleaned and re-used the same substrate without a noticeable degradation of its cleanliness. In addition, we have a few experimental results on Au on sapphire substrates that are in agreement with the results shown here. We first clean the Au(111) surface by Ar ion sputtering and annealing cycles in our vacuum chamber with a base pressure of $10^{-9}$ Pa. For direct current heating we mounted the Au on mica substrate on a Ta stripe for direct current heating. We then checked for each sample that the herringbone reconstruction of Au with no signs of contamination was observed using SFM (VT-AFM-XA from Omicron, Taunusstein, Germany) in the same vacuum vessel. We deposited 10,10'-Dibromo-9,9'-bianthryl (DBBA) on the substrate pre-heated to $180^{\circ}$ C. After that we increased the substrate temperature to $250^{\circ}$ C for $2$ h followed by another annealing step at $420^{\circ}$ C for $1$ h. During this process graphene nanoribbons form due to on-surface Ullmann coupling. More details on the nanoribbon formation process can be found in \cite{schneider16p1,bytyqi18u1} where we also describe our investigations of the graphene nanoribbon formation process as a function of sample annealing temperature. After molecule deposition and annealing, the sample is again transferred to the SFM cooled down to about $115$ K using liquid nitrogen. Imaging was performed about 40 h after the cleaning of the Au substrate.

\subsection{Scanning Force Microscopy methods}

We use Si cantilevers (Nanosensors) with a longitudinal force constant of $c_{\mbox{\scriptsize L}} \approx 4$ N/m, a fundamental eigenfrequency of $f_1 \approx 80$~kHz and a pyramidal tip. The low force constant and fundamental eigenfrequency are needed because the photo-detector has a cut-off frequency of about $500$ kHz and the second resonance frequency occurs at about 6 times the fundamental resonance frequency. To oscillate the tip and to detect the first and second frequency shift we use two phase-locked loop systems (NANONIS, Specs, Switzerland). The distance of the tip to the surface is controlled by a feed-back loop operating on the frequency shift of the fundamental eigenfrequency. We cleaned the tip by Ar ion sputtering. Since the tip remains at room temperature in our setup while the sample is cooled down to $115$ K, we did not attempt to functionalize the tip. Internal temperature gradients lead to a considerable thermal drift, but the results shown below are reproducible for several images within the experimental uncertainty. Backward and forward scans give similar results. The contact potential difference between tip and sample, often on the order of $1.0$ to $1.5$ V, is compensated during imaging by applying a constant voltage to the tip. In spite of this compensation, we expect a charge transfer between Au and the graphene nanoribbon and the occurrence of local electrostatic forces as discussed in detail below.

\subsection{Bimodal measurement mode}

The frequency shift in the first mode without bimodal excitation is given by \cite{giessibl00p1}

\begin{equation}
 \Delta f_1=\frac{f_1}{2\pi c_{\mbox{\scriptsize L,1}}}\int_0^{2\pi}F\left[z_0 + A_1\cos\theta_1\right]\cos\theta_1\mathrm{d}\theta_1
\end{equation}

whereas the frequency shift in the second mode with bimodal excitation \cite{kawai09p1}

\begin{equation}
 \Delta f_2=-\frac{f_2}{2\pi c_{\mbox{\scriptsize L,2}}A_1}\int_0^{2\pi}F'\left[z_0 + A_1\cos\theta_1\right]\cos\theta_1\mathrm{d}\theta_1
\end{equation}

where $F(z)$ is the tip-sample force as a function of distance and $F'=\mathrm{d}F/\mathrm{d}z$, $A_1$ and $A_2$ are the oscillation amplitudes of first and second mode and $\theta_1$ is the phase of the oscillation in the first flexural mode. Since $A_1\gg A_2$ and $A_2$ is small with respect to changes in $F(z)$, the measured frequency shift $\Delta f_2$ is related to the first derivative of the force with respect to $z$, an integral over $A_2$ is not needed and the parameter $A_2$ does not appear in the formula. Both the derivative and the frequency shift generally show a stronger distance dependence than the original force. Also for measurements without bimodal oscillation, the frequency shift $\Delta f$ is related to the derivative of the force, if small oscillation amplitudes are used. For bimodal measurements, however, the integral is taken over the full oscillation cycle of the oscillation at the fundamental eigenfrequency with a large oscillation amplitude $A_1$. Over a large part of the oscillation cycle the contribution of the short-range force derivative is near zero, weakening the overall intramolecular contrast, but enhancing the sensitivity to long-range forces, as for example electrostatic contributions.

\subsection{Force curve measurement and fitting}

Frequency shift has been measured as a function of distance above a graphene nanoribbon and above the Au(111) surface. The tip-sample distance of the frequency shift above the graphene nanoribbon indicates a frequency shift of $-44.5$ Hz used during imaging at $z=0$ as expected. The frequency shift data measured above Au(111) (red data points) was adjusted such that at a frequency shift of $-44.5$~Hz the distance between the two curves was $-0.065$ nm as expected from imaging. The frequency shift was then converted to force using Baratoff's method briefly described in ref \cite{pfeiffer02p1}. The resulting total force was fitted using functions that had been used in previous publications to describe van-der-Waals and electrostatic forces of a conical tip with a spherical cap.

The results show reasonable agreement, but the fit parameters gave a tip radius of 15 to 30 nm - unusually large given the values provided by the manufacturer (tip radius below 8 nm) and considerably larger than previous results. We therefore abandoned the van-der-Waals model and neglected van-der-Waals forces in order to reduce the number of fit parameters. Since charge-transfer is expected for the graphene on Au(111) system \cite{vanin10p1}, we used appropriate electrostatic force models as described in the text.

\begin{figure}[t]
\centerline{
\includegraphics[height=10cm]{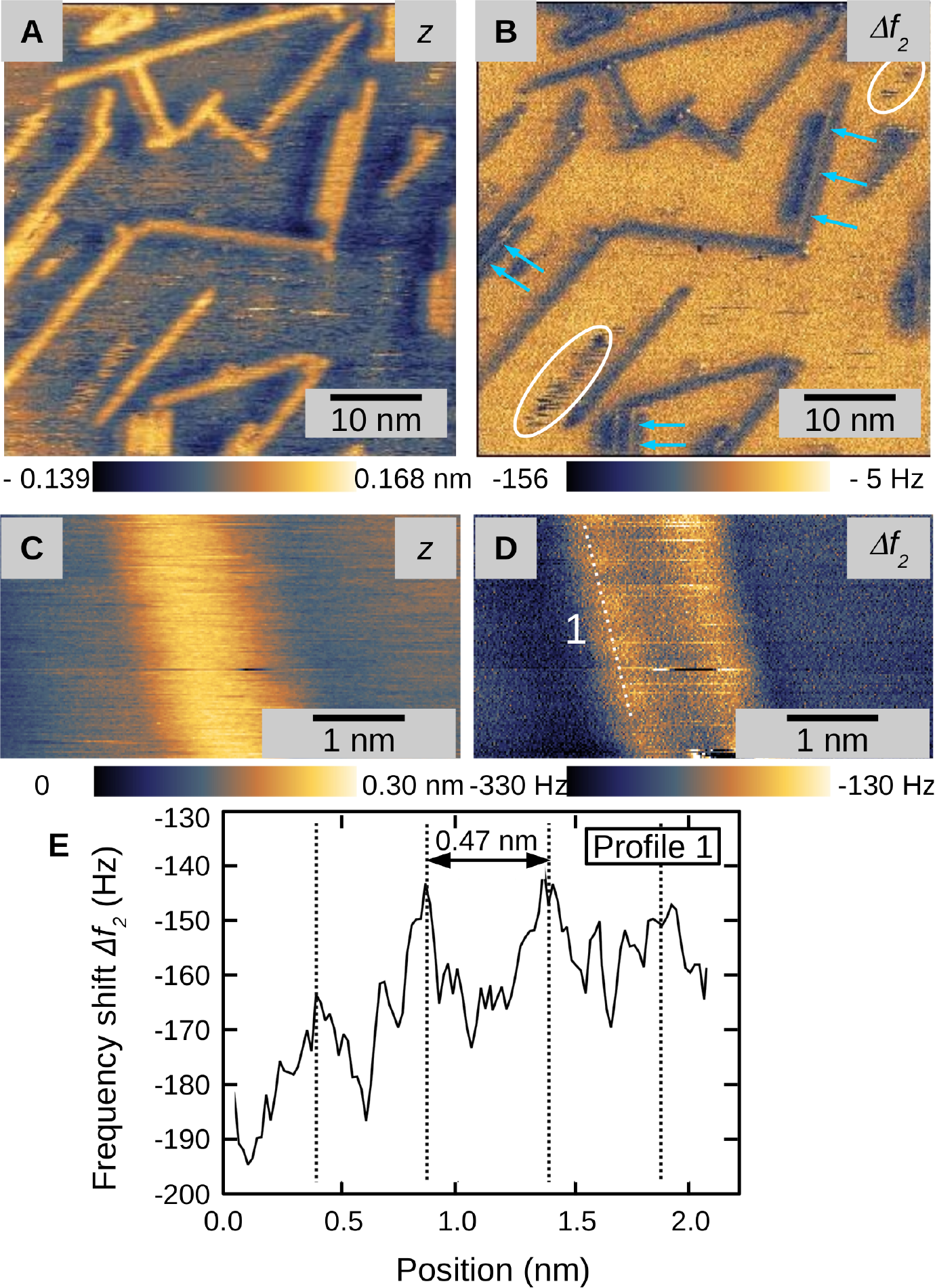}}
    \label{fig1}
\end{figure}

\paragraph*{Fig. 2: Scanning force microscopy images of graphene nanoribbons using bimodal oscillation.} ({\bf A}) Topography and ({\bf B}) corresponding image of the second frequency shift $\Delta f_2$ showing an overview with several graphene nanoribbons. Bright lines at the edge of graphene nanoribbons have been marked with blue arrows. Mobile graphene nanoribbons have been marked with white ovals. $\Delta f_1 = - 44.5$~Hz, $A_1 = 15$~nm, $A_2 = 50$~pm, $c_{\mbox{\scriptsize L}} = 4.0$ N/m, $f_1 = 78$ kHz, $f_2 = 496$ kHz. ({\bf C}) Topography and ({\bf D}) corresponding bimodal image of $\Delta f_2$ showing a single graphene nanoribbon. The nanoribbon-edges show an internal structure. $\Delta f_1 = - 50$~Hz, $A_1 = 15$~nm, $A_2 = 50$ pm, $c_{\mbox{\scriptsize L}} = 4.0$ N/m, $f_1 = 78$~kHz, $f_2 = 496$~kHz. ({\bf E}) Frequency profile No 1 obtained after averaging over $0.2$ nm width along the dotted white line in D.

\section{Results and discussion}

\subsection{Bimodal images - overview}

In Figure \ref{fig1}A we show a typical topography image in bimodal mode with the feed-back loop acting on the shift of the fundamental eigenfrequency $f_1$. Compared to the structures formed during previous steps of the on-surface reaction, the graphene nanoribbons appear flat and show little internal structure in overview images. In Fig. \ref{fig1}B we show the frequency shift $\Delta f_2$ measured with respect to the second harmonic. $\Delta f_2$ images (Fig. \ref{fig1}B) mostly reflect the topography (Fig. \ref{fig1}A), but the graphene nanoribbons appear dark compared to the Au surface. In areas marked by blue arrows the graphene nanoribbons show bright lines at the edges. These lines are mainly visible between two neighboring graphene nanoribbons. Mobile graphene nanoribbons have been marked with white ovals.

Figures \ref{fig1}C and D show an enlarged region of a single graphene nanoribbon. While in the topography in Figure \ref{fig1}C no intramolecular resolution is achieved, the $\Delta f_2$ measurement shows the onset of intramolecular resolution, serrated edges of the nanoribbon. A $\Delta f_2$ profile of the same area along the dotted line is shown in Figure \ref{fig1}E. The maxima have a distance of about $0.47$~nm averaged over several periods and result from the armchair configuration of the graphene nanoribbon edge for which a distance of 426 pm which corresponds to 3 C-C distances, 142 pm,  is expected\cite{harrison80p1}.

\newpage
\begin{figure}[h]
\centerline{
\includegraphics[height=10cm]{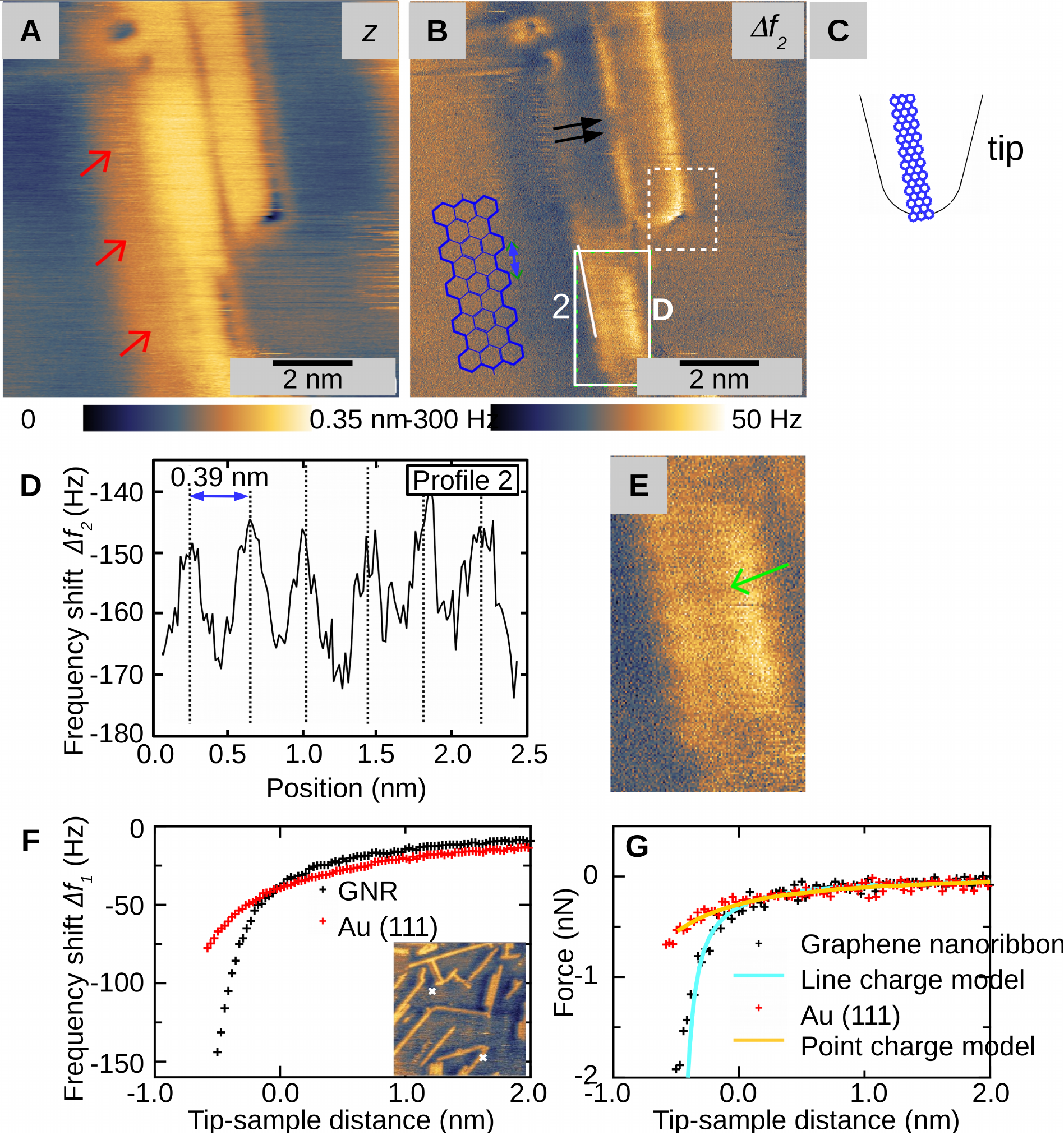}}
    \label{fig2}
\end{figure}

\paragraph*{Fig. 3: Bimodal scanning force microscopy images of two adjacent graphene nanoribbons.}
({\bf A}) Topography and ({\bf B}) corresponding bimodal image of the second frequency shift $\Delta f_2$. Two missing bright features at the nanoribbon edge are indicated by two black arrows.  $\Delta f_1 = - 56$~Hz, $A_1 = 15$ nm, $A_2 = 50$~pm, $c_{\mbox{\scriptsize L}} = 4.0$~N/m, $f_1 = 78$ kHz, $f_2 = 496$ kHz, scanning speed $13$ nm/s.({\bf C}) Model of the tip. ({\bf D}) Frequency shift $\Delta f_2$ profile averaged over $0.2$ nm width along the white line in B numbered 2. ({\bf E}) Enlarged area cut from image B at the position of the white rectangle. ({\bf F}) Frequency versus distance measurements taken at the positions shown in the inset. The relative tip-sample distance of the two datasets has been adjusted. ({\bf G}) Force data obtained by conversion from the frequency shift data.

\newpage
In Figure \ref{fig2}A, B taken with the same macroscopic tip, we show the topography and $\Delta f_2$ of two adjacent graphene nanoribbons. In Figure \ref{fig2}A a red arrow marks an area located next to the graphene nanoribbon where a  bright area is observed. Also the graphene nanoribbon on the left hand side appears brighter due to the vicinity of the adjacent graphene nanoribbon. This bright area is caused by the shape of the tip and results in a different shadow-like $\Delta f_2$ of the two structures in Figure \ref{fig2}B. Such features are also visible in the overview images Figures \ref{fig1}A and B, but become hard to see for part of the orientation angles. The size of the shadow points to a graphene nanoribbon absorbed on the tip, see Figure \ref{fig2}C and Supplementary Information\dag, section 2. The graphene nanoribbon on the tip exposes a tilted zigzag edge to the sample with one corner of the zigzag edge forming the tip apex. In other related experiments, we have manipulated graphene nanoribbons on the surface \cite{schneider16p1}. In the Supplementary Information\dag, Section 1, an image without a graphene nanoribbon adsorbed to the tip is shown for comparison and images with tip changes are shown in section 3. In such images, also bright edges are observed. In previous experiments where the tip is cooled in contrast to our work, the tip was deliberately functionalized using a CO molecule\cite{dienel15p1} or with CuO, see ref \cite{moenig16p1}. Since the tip remains at room temperature in our SFM, functionalizing the tip with a graphene nanoribbon allows to obtain a robust, sharp and reproducible tip suitable for high-resolution imaging.

Figure \ref{fig2}B shows, similar to Figure \ref{fig1}D, serrated edges as becomes clear also in the line cut shown in Figure \ref{fig2}C. The serrated edges are best visible on the graphene nanoribbon on the right hand side, but they are also present on the graphene nanoribbon on the left hand side in the upper and lower region and are even visible in Figure \ref{fig2}A as dark spots arranged in the form of a line. The part marked by a white rectangle is shown in large in Figure \ref{fig2}E. For comparison, a schematic drawing of the structure of the graphene nanoribbon on the sample has been added in blue (not to scale). The maxima in Figure \ref{fig2}D have a distance of about $0.39$~nm averaged over several periods. The region of the strongest frequency shift is observed at the short edge of the graphene nanoribbon near its corner and has been marked by a white dotted rectangle. In the upper third of Figure 2B, on the left edge of the graphene nanoribbon located on the right hand side, there are two bright features missing, indicated by two black arrows. This defect has been imaged in several consecutive images. The observation of atomic-scale defects, not only intramolecular periodicity, shows the true high resolution of the measurement in contrast to tip-induced averaging.

In contrast to previous results\cite{dienel15p1}, in our measurements the edges of the nanoribbons appear brighter than their center, in Figure \ref{fig1}D and in Figure \ref{fig2}B and E. This result is reproduced with several tips and separate samples when they are cooled to 115 K. We attribute this observation to an additional repulsive interaction at the edges. This force could be caused by electrostatic contributions and/or by the Pauli repulsion. It seems natural that the graphene nanoribbon on the tip should be terminated by the same atomic species as the one on the surface and the repulsive interaction results from the interaction of similar atomic species. Images with other tip terminations are shown in the supplementary information, section 3. For other tip terminations, the interaction at the edges is attractive. This points to electrostatic forces at the graphene nanoribbon edges, because electrostatic forces can be attractive or repulsive depending on the nature of an electrostatic charge located at tip apex \cite{foster01p1}.

In previous results graphene nanoribbons have been shown to be hydrogen-terminated. However, hydrogen termination is not expected to show large repulsive interaction, and usually the interaction with hydrogen is not large enough to detect hydrogen in scanning force microscopy. We therefore tentatively suggest that our graphene nanoribbons are not hydrogen-terminated. If hydrogen were absent here, leaving unterminated carbon, we would expect adjacent graphene nanoribbons to merge. In contrast we often observe a bright line between adjacent graphene nanoribbons (blue arrows in Figures \ref{fig1}B and a bright line can also be seen in \ref{fig2}B) and conclude that the edges must be atomically protected. We therefore consider that the graphene nanoribbons could be metal-terminated. Au surface steps are known to be mobile even at room temperature and Au surface diffusion during sample annealing and cool-down could lead to Au trapping at the graphene nanoribbon edges.

One might suppose that the bright egdes could be caused by elastic deformations at hydrogen-terminated graphene nanoribbon edges or from the interaction of a hydrogen-terminated graphene nanoribbon with the Au herringbone reconstruction. Here, we exclude these possibilities, because they have not been observed in previous publications, e.g. \cite{dienel15p1}. We do take into account that the local topography as well as elastic and inelastic deformations could differ from previous results due to Au trapping at the graphene nanoribbons edges.

\subsection{Force-distance data - electrostatic interactions}

The frequency shift versus distance has been measured at the positions indicated in Figure \ref{fig2}F above a graphene nanoribbon's edge and above the Au(111) surface. The force calculated from the frequency-distance measurement is shown in Figure \ref{fig2}G together with a model. For the force measured above the Au(111) surface we have used a point charge on the tip at the graphene nanoribbon corner interacting with its own image charge located in the Au(111) surface. The only free parameters of the model are the charge, which is obtained to be 1.85 $e$ and an offset with respect to the experimentally measured distance such that the point of divergence of the model occurs at $-1.7$ nm.

For the force observed on the graphene nanoribbon we used a line charge model, i.e. we considered the electrostatic field of a charged rod given by

\begin{equation}
E=\frac{\lambda}{4\pi\varepsilon_0}\left(\frac{\alpha}{r_1}-\frac{\beta}{r_2}\right)
\end{equation}

where $\lambda$ is the charge per unit length, $\varepsilon_0$ is the permittivity constant, $\alpha$ and $\beta$ are constants of order 1 accounting for the orientation and positioning of the rod with respect to the point of the measurement. $r_1$ and $r_2$ are the respective distances between the point of measurement and the ends of the rod. The two edges of the graphene nanoribbons could give two such contributions and the charged corner could give a point-like contribution in addition. In order to simplify the problem and reduce the number of fit parameters, we used as a test function for the force

\begin{equation}
F(z)=\frac{\alpha\cdot n e\cdot\lambda}{4\pi\varepsilon_0}\,\cdot\,\frac{1}{z}
\end{equation}

where $e$ is the elementary charge, $n$ is the number of charges on the tip and $z$ is the tip-sample distance representing the interaction between the point-like charge of the tip graphene nanoribbon corner and a line-like graphene nanoribbon on the surface. The result is an excellent fit providing $\alpha\cdot n\cdot\lambda$ equal to 0.656 $e$/nm. Fitting was performed for tip-sample distances above $-0.2$ nm, but the resulting curve represents the data well also at smaller tip-sample separations.

Setting $\alpha=1$ and assuming $n=1.85$ (the number of charges on the tip) from the fit of the force measured above Au, we obtain $\lambda$ equal to 0.35 $e$/nm. This translates into a charge of 0.075 $e$ per atom for each of the two atoms in the edge unit cell of length $426$ pm. This charge per unit length is a reasonable value at the scale of a graphene nanoribbon: Density functional theory including the van-der-Waals interaction has predicted an overall charge transfer of $+0.04\,e$ per atom (p-doping) from the Au surface \cite{vanin10p1}.

\begin{figure}[h]
\centerline{
\includegraphics[height=4.3cm]{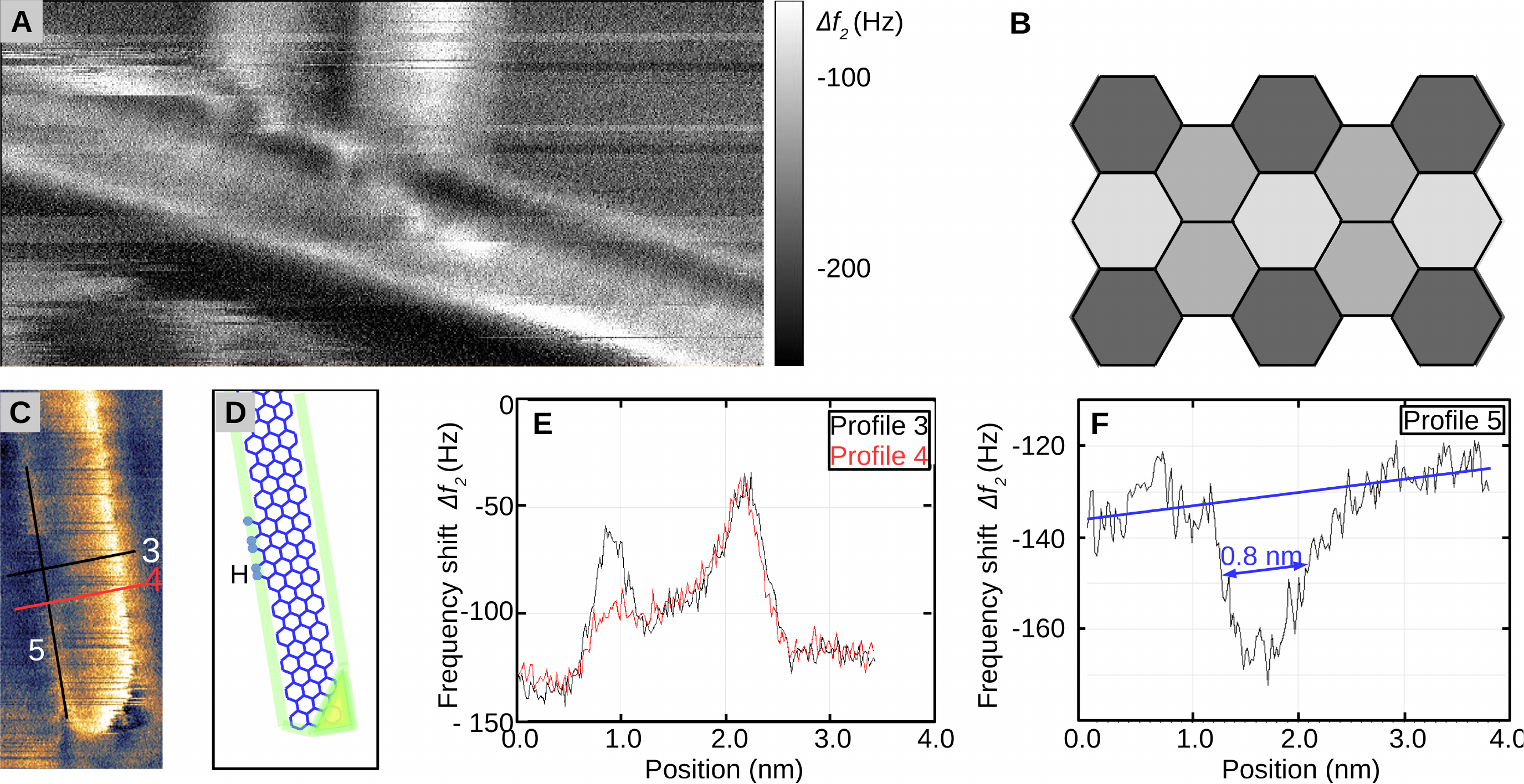}}
    \label{fig4}
\end{figure}

From the overall charge of 1.85 $e$ at the tip we obtain a charge of 0.185 $e$ per atom if we assume that this charge is distributed over 10 atoms. We estimate this size from the size of the area strongly affected by additional forces in Figure \ref{fig4}C. This value is about $2.5$ times larger than the one observed on the body of the graphene nanoribbon. Charge enhancement factors of up to 10 are predicted to occur due to electrostatic effects of the extended Dirac fermion system \cite{wang10p1}. For the length of many graphene nanoribbons of 10-20 nm and their width of about 1.5 nm, a charge enhancement of about 2.5 is predicted \cite{wang10p1}. The overall agreement of this value with the result from our force curve fitting procedure supports our analysis of the electrostatic forces in this system. In our measurements, we observe a clear difference between the two corners of the graphene nanoribbon, in simulations the two corners are equal. We attribute this difference to the presence of the adjacent nanoribbon.

Independently of the electrostatic model used, the force measured above the graphene nanoribbon is larger than the one measured on the Au surface. This indicates that the charge on the tip must have a different sign compared to the charge on the graphene nanoribbon on the surface. We suggest that the difference of charge sign arises because the tip exposes an {\it end} of the graphene nanoribbon, while on the surface we measure above the {\it center} of a graphene nanoribbon. This could be explained by the interplay between charge transfer and charge enhancement at the graphene nanoribbon corners. Another possibility is that the tip graphene nanoribbon interacts with a different substrate compared t Au, possibly Si and shows a different charge transfer. In addition the charge distribution could be influenced by the Smoluchowski effect \cite{ellner16p1}.

The effect of screening by the Au surface is surprisingly small. The reason for the weak screening could be that the tip approaches to a rather close distance to the graphene nanoribbon, while the Au substrate under the graphene nanoribbon remains at a distance to the tip of several Angstr\"oms, and the charge induced in Au by screening is located at an even larger distance to the tip below the Au surface. The shape of the force distance curve is not strongly altered by atomic relaxation, probably due to the well-known rigidity of graphene.

In the force-distance data we expect an exponential attractive contribution for covalent bonding interactions at small tip-sample distances. We cannot detect such an attractive force in these force-distance data. Below we discuss that chemical interactions related to Clar's rule do play a role on the body of the graphene nanoribbon. At the smallest tip-sample distances we observe a repulsive deviation from the model due to repulsive short-range forces. We conclude that even though chemical interactions are expected to be present, they are not the dominating interaction. This is in line with our model where the tip graphene nanoribbon's edge is atomically protected by metal.

\paragraph{Fig. 4: Bimodal scanning force microscopy image and line-cuts through defective area.} ({\bf A}) Image showing stripe-like contrast. $\Delta f_1 = - 71$~Hz, $A_1 = 15$ nm, $A_2 = 50$ pm, $c_{\mbox{\scriptsize L}} = 4.0$~N/m, $f_1 = 78$ kHz, $f_2 = 496$~kHz, scanning speed $13$ nm/s ({\bf B}) Contrast of the graphene nanoribbon expected from Clar's considerations. Light gray benzene rings are expected to be chemically more reactive compared to dark gray ones. ({\bf C}) Another $\Delta f_2$ image of the same graphene nanoribbon as in Figure 2. ({\bf D}) Schematic of the graphene nanoribbon including the defective region drawn to the scale of the image A. ({\bf E}) Frequency shift $\Delta f_2$ profiles averaged over $0.4$~nm width along the black and red lines in part C numbered 3 and 4. ({\bf F}) Line cut taken from the black line indicated in part A and numbered 5. The line cut was averaged over a width of $0.2$~nm. $\Delta f_1 = - 65$~Hz, $A_1 = 15$ nm, $A_2 = 50$~pm, $c_{\mbox{\scriptsize L}} = 4.0$ N/m, $f_1 = 78$ kHz, $f_2 = 496$~kHz, scanning speed $15$~nm/s.

\newpage

\subsection{Stripe-like contrast}

In Figure \ref{fig2}E, we observe a weak stripe-like contrast in the center of the graphene nanoribbon and have marked it by a green arrow. This contrast is not induced by tip changes and occurs at a small angle with respect to the fast scan direction. Such a stripe-like contrast is also observed in additional images, Figure \ref{fig4}A, in a direction nearly perpendicular to the fast scan direction. A charge modulation or local differences of the chemical reactivity due to Clar's rule \cite{dienel15p1,martin12p1} could explain this contrast, see Figure \ref{fig4}B.

\subsection{Defective area at the edge}

Figure \ref{fig4}C shows an enlarged view of the nanoribbon shown on the right hand side in Figure~ \ref{fig2}B with enhanced contrast taken from a subsequent image. Two line-cuts across the graphene nanoribbon, indicated by lines numbered 3 and 4 in Figure \ref{fig4} C, are shown in Figure \ref{fig4}E. One line-cut is taken at the position where a bright feature is observed and the other is taken at the position where it is missing. A line cut along the edge of the graphene nanoribbon through the defective area allows to estimate the number of defective sites from the size of the dip in the contrast, see Figure \ref{fig4}F. The dip is about $0.8$ nm long, measuring its length at half of its depth. This length corresponds to twice the distance between anthracene units, $0.4$ nm, and to four atomic sites. Since the left hand side of the dip (this side corresponds to the upper side of the nanoribbon in Figure \ref{fig4}C is much steeper than its right hand side (the lower side of the nanoribbon Figure \ref{fig4}C), there could be an additional defective atomic cite such that there are five in total.

Although it is expected that hydrogen atoms cover the edges, we re-assess this point due to the different contrast we observe in our images compared to literature and suggest that the edges are metal-covered.  In this model the defective area could be formed by five hydrogen atoms replacing metal atoms. In the Supporting Information of ref \cite{ruffieux16p1}, bright contrast in scanning force microscopy images on a graphene nanoribbon with long zigzag edges has been attributed to missing hydrogen. In scanning tunneling microscopy images, the authors associate hydrogen defects with bright contrast. The authors also show that tip-induced manipulation removes the defective site. In Ref. \cite{ao10p1} the enhanced stability of the graphene-graphane interface of graphene nanoribbons is discussed. Such an enhanced stability could be important for the interface of the defect observed here with the graphene nanoribbon edge. In Ref. \cite{chen15p1} the authors find an additional density of states on the edges of graphene nanoribbons by scanning tunneling microscopy measurements. This density of states at the edge extends further into the vacuum than the states in the center of the graphene nanoribbon.

From the process of Ullmann coupling we expect that hydrogen is removed from the graphene nanoribbon at 420 K. The abundance of hydrogen on the surface at the time of observation could depend on the details of the experiment. In most of the cited work, the authors investigate graphene nanoribbons placed inside bath cryostates at temperatures as low as 4-10 K. These temperatures are below the sublimation point of hydrogen that could be trapped inside the cryostate. The graphene nanoribbons could thus be re-hydrogenated upon cool-down. In this work we cool down only to 115 K in a flow cryostate while the tip remains at room temperature. Although we expect that also at 115 K hydrogen should eventually cover the graphene nanoribbon edges, we suggest that this has not yet occurred at the time of observation.

\section{Conclusions}
We use a bimodal imaging technique exciting the first and second eigenmode of a cantilever to investigate graphene nanoribbons cooled by a flow cryostate at liquid nitrogen temperatures with the tip at room temperature. Force-distance measurements allow us to understand and quantify the electrostatic forces above the graphene nanoribbons. We also show that our graphene nanoribbons are probably hydrogen-free with the exception of a small defective area modelled by five hydrogen atoms. We suggest to dose the graphene nanoribbons with hydrogen in further experiments to investigate the magnetic properties of partly hydrogenated graphene nanoribbons.

\section*{Acknowledgements}
We thank K. Bytyqi, M. Marz, C. P\'erez Le\'on and O. Stetsovych for help with the experimental work and F. Pauly, G. Xing, E. Scheer, H. v. L\"ohneysen, R. Danneau and S. Diesch for stimulating discussions. This work was supported by the ERC Starting Grant NANOCONTACTS (No. 239838).
\newpage

\section{Supporting Information: On-surface synthesis of graphene nanoribbons from DBBA oligomers and images obtained at room temperature}

On-surface synthesis of the graphene nanoribbons from 10,10'-dibromo-9,9'-bianthryl (DBBA) (ref \cite{cai10p1,schneider16p1,bytyqi18u1}) relies on the formation of DBBA oligomers. In the DBBA oligomers there are still hydrogen atoms present protecting the oligomers from forming graphene nanoribbons. Further heating allows to remove the hydrogen atoms and form graphene nanoribbons.

Large-amplitude scanning force microscopy images (Fig. S\ref{figS2}) show a clear height difference between DBBA oligomers and graphene nanoribbons - demonstrating that the final synthesis step where the hydrogen atoms are removed from the DBBA oligomer is successful. DBBA oligomers are detected from their increased height, from the zigzag shape of the alternating anthracene units and from the period of their zigzag structure whereas the graphene nanoribbons show no internal structure in standard large-amplitude scanning force microscopy images. The width ($0.14$ nm) and height ($0.19$ nm) of the graphene nanoribbons can be measured from the line profile (Fig. S\ref{figS2}D) and compared to the height of a DDBA oligomer ($0.30$ nm). For other measurements, the height of graphene nanoribbons appears even smaller ($0.8$ nm). The period of the zigzag structure observed on the DBBA oligomers is expected to be approximately $0.85$~nm - twice the value measured for the internal structure of graphene nanoribbons. 

\begin{figure}[h]
\centerline{
\includegraphics[width=9cm]{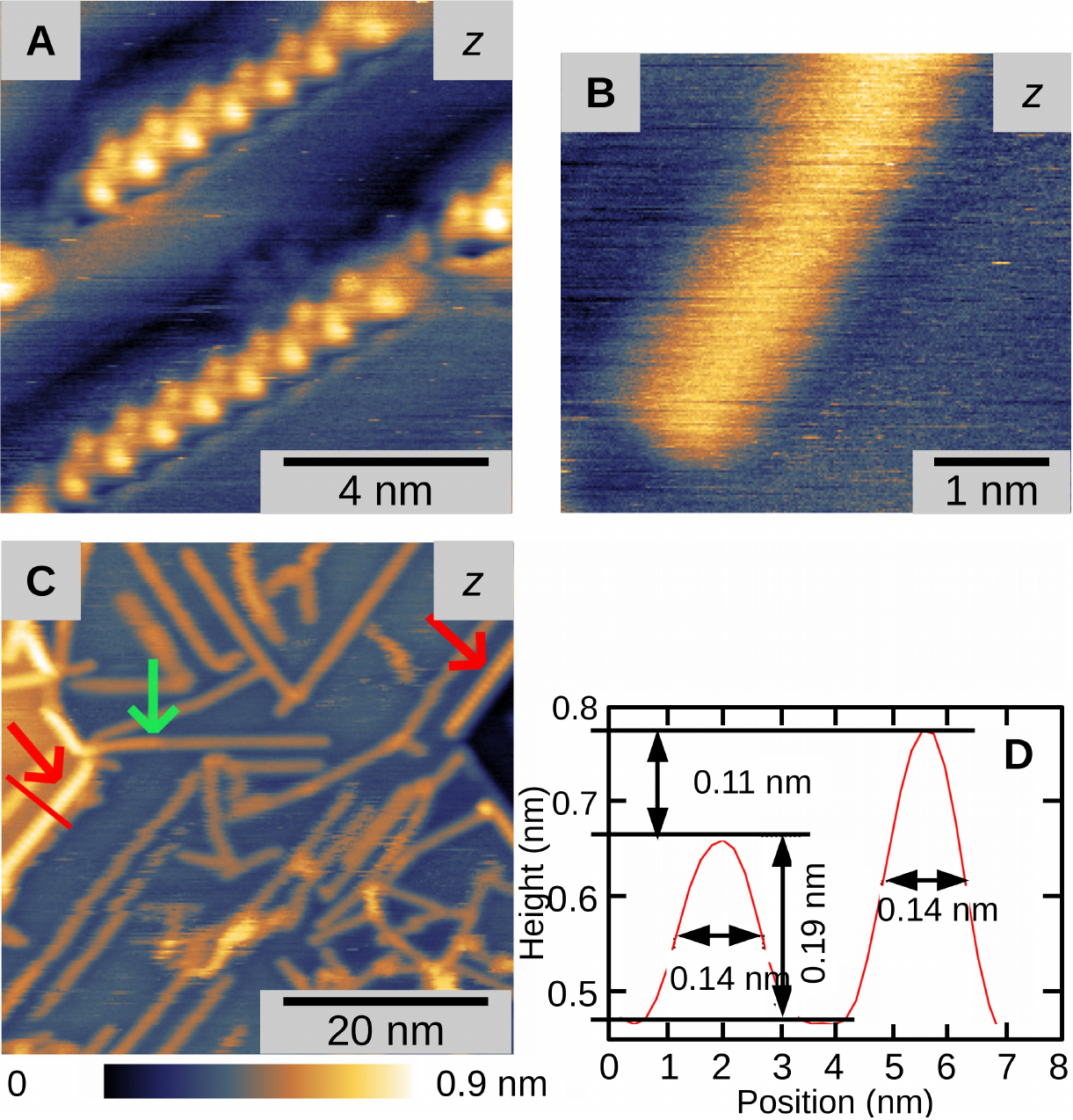}}
    \label{figS1}
\end{figure}

\paragraph*{Fig. S1: Images using standard large-amplitude cantilever SFM.} ({\bf A}) Image of a DBBA oligomer. The DBBA oligomers clearly show a zigzag internal structure easily resolved by SFM. The period is approximately 1 nm in this image. $\Delta f=-27$ Hz, $A=8$ nm, $c_L=43$ N/m, $f_1=351$ kHz. ({\bf B}) Image of a graphene nanoribbon. 
    ({\bf C}) Overview image showing graphene nanoribbons on a Au(111) surface after on-surface synthesis. The red arrows show DBBA oligomers. The green arrow shows a junction between a graphene nanoribbon and a DBBA oligomer. The profile shown in B is taken at the position of the red line. ({\bf D}) Profile along the red line in C showing a direct comparison of a DBBA oligomer with a graphene nanoribbon. The DBBA oligomer is $0.11$ nm higher than the graphene nanoribbon. $\Delta f=-29$ Hz, $A=7$~nm, $c_L=40$ N/m, $f_1=340$ kHz.

\newpage

\begin{figure}[t]
\centerline{
\includegraphics[width=5cm]{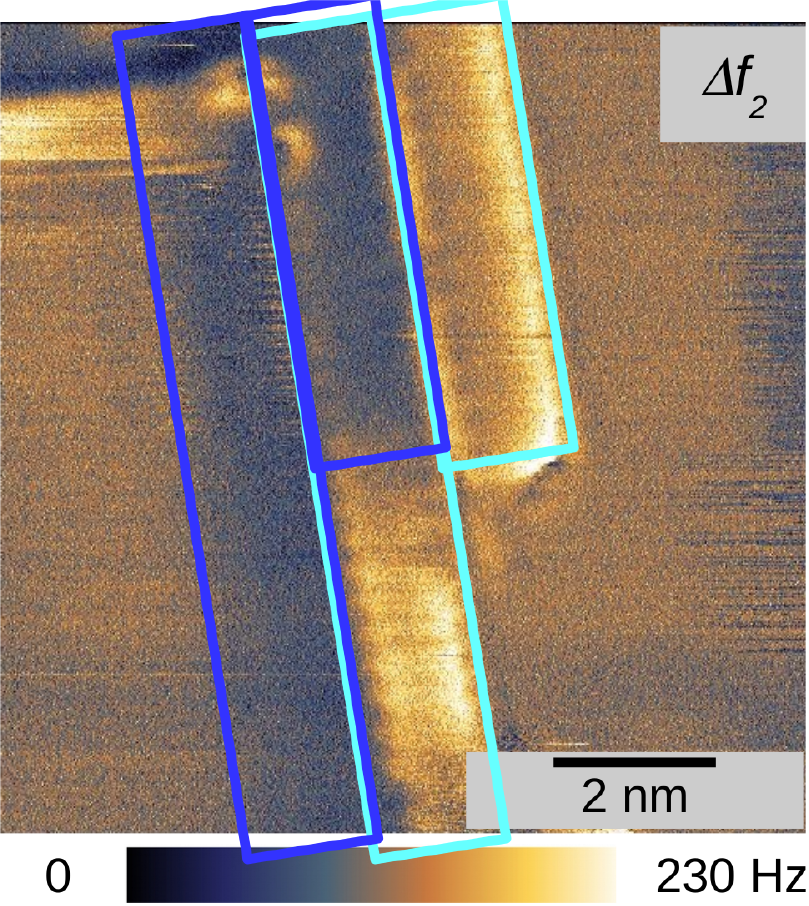}}
    \label{figS2}
\end{figure}

\paragraph*{Fig. S2: Indication of the shadow of the two nanoribbons depicted in dark blue.} The two dark blue rectangles are the same as the two light blue rectangles indicating the brighter graphene nanoribbons.

\section{Supporting Information: Tip shape and termination}

In Figure 2a of the main text, a bright shadow is observed next to the two adjacent nanoribbons and has been marked by red arrows. This shadow in Figure 2A corresponds to a darker region at the same position relative to the nanoribbons in Figure 2B. The darker area is about $35$ Hz lower in frequency shift compared to the adjacent Au terrace translating to a distance of $62$ pm. The size and shape of the shadow is similar in forward and backward scans and remains similar when changing the scanning speed. Figures~2 and 3 have been acquired using different scanning speeds while maintaining the same image size.

The size and shape of the shadow is precisely that of the longer graphene nanoribbon observed in the center of the image and adjacent to the right hand side of the shadow. In addition, there is a shadow overlaid on the upper part of the graphene nanoribbon located at the left hand side also about $35$ Hz lower compared to the brighter part of the nanoribbon. This shadow has precisely the shape of the shorter graphene nanoribbon observed on the right hand side of the image and adjacent to the right hand side of the shadow. We indicate the two shadows by two dark blue rectangles on Figure S2. The shorter nanoribbon appears somewhat wider than the longer one and its edge covers the edge of the longer nanoribbon. Maintaining the same distance between the two light blue rectangles and moving them sideways by precisely the width of one graphene nanoribbon generates the two dark blue rectangles in registry with the shadow.

These observations point to a double tip where the distance of the two tips is precisely the width of a graphene nanoribbon. Since the two edges of the graphene nanoribbons give the strongest interaction, we argue that a graphene nanoribbon has been adsorbed to the tip, see model in Figure S3. In this model, for the tip we have assumed a radius of $2$~nm in accordance to the values given by the manufacturer and in agreement to a quantitative description of long-range electrostatic forces in scanning force microscopy measurements, as has been given e.g. in ref \cite{lantz03p1}. One of the corners of the graphene nanoribbon gives the strongest tip-sample interaction and generates the main image. The other corner is located at a slightly larger distance from the sample surface and merely generates a dark shadow overlaid to the main image.

We can even determine the orientation of the graphene nanoribbon. Since the shadow next to the graphene nanoribbons on the surface has precisely the width of a graphene nanoribbon, the graphene nanoribbon on the tip must be oriented parallel to the graphene nanoribbons on the surface in this image. In this configuration it exposes a zigzag edge to the sample. One of the two corners of the zigzag edge gives a brighter image, because the zigzag edge is tilted with respect to the surface. In this configuration one corner of the zigzag edge forms the tip apex.

\newpage
\begin{figure}[h]
\centerline{
\includegraphics[width=8cm]{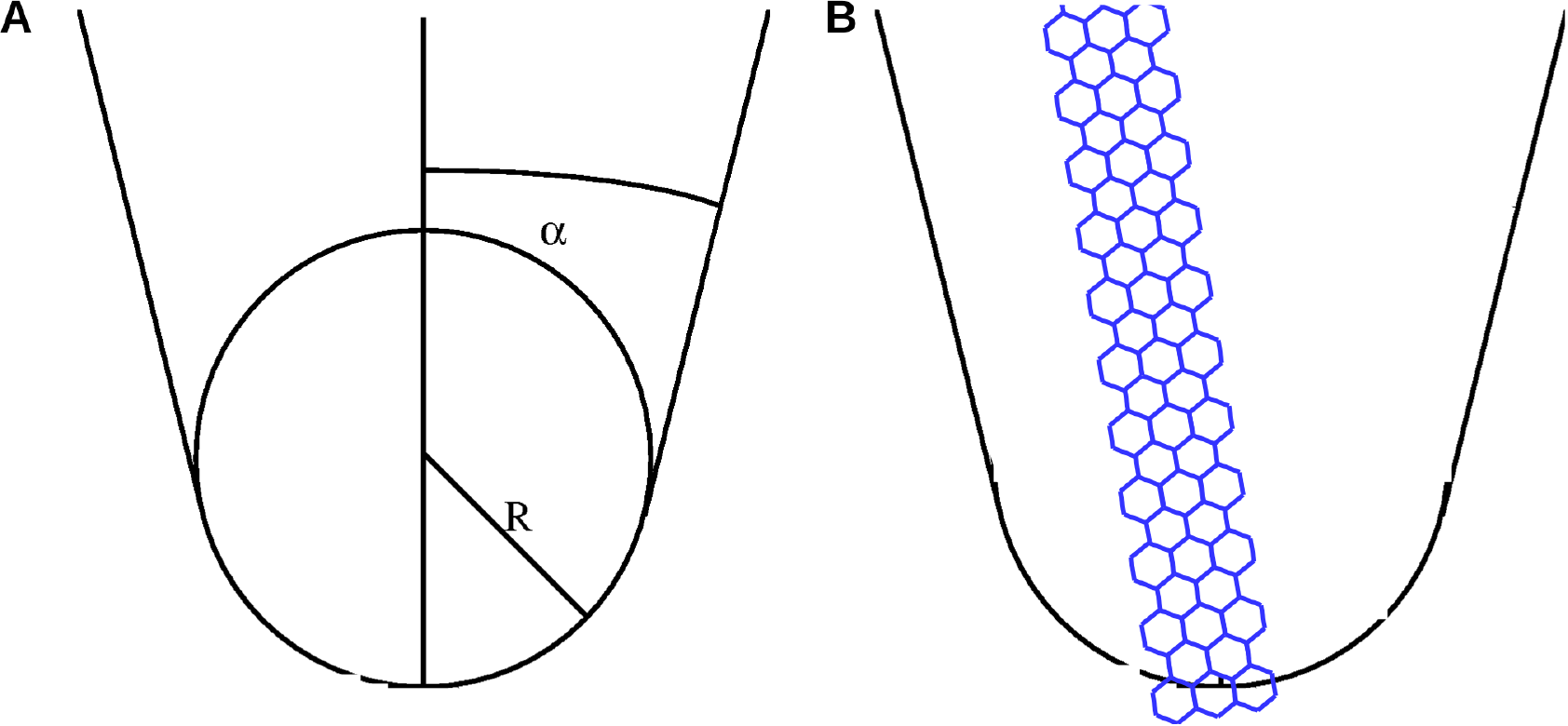}}
    \label{figS3}
\end{figure}

\paragraph*{Fig. S3: Model of the tip.} ({\bf A}) Usually the manufacturer supplies an opening angle and a tip radius. ({\bf B}) We assume a tip radius of $2$~nm, the nanoribbon, depicted in blue, has a width of about $1$~nm and a thickness of several \AA.

\newpage
\section{Supporting Information: Bright edges of graphene nanoribbons observed through tip change at 115 K}

In the main article we discuss how bimodal mode enhances the contrast at the edges of the graphene nanoribbons and the sensitivity of the measurement to long-range electrostatic forces. In these measurements, the contrast in the fundamental mode topography images remains as well-known for graphene nanoribbons. In this section, we show images obtained in monomode dynamic frequency modulation scanning force microscopy images where the contrast has changed by a tip change. In these images in the topography channel the edges also look bright. In these measurements the tip has been cleaned by Ar ion sputtering.

To obtain the images shown here, first we imaged a region containing graphene nanoribbons using scanning force microscopy (Fig. S4A). Then a disordered region of the surface was imaged using the same tip. Subsequently, we imaged again the same region containing graphene nanoribbons as before. The tip has changed (Fig. S4B) and after the tip change bright contrast is observed on the edges of the graphene nanoribbons. When we took another image at the same area, the tip had changed back to its original state (Fig. S4C). For Figure S4B we have no indications that a graphene nanoribbon could have been adsorbed to the tip.

We compare these measurements to Fig. 3A of the main text. In the two different measurements the tip-sample distance could differ and the relative importance of different types of forces could be different. In particular the role of the chemical interaction should strongly depend on the chemical details of the tip apex. It is remarkable that in Fig. S4B the interaction of the tip and the graphene nanoribbon's edge is attractive in contrast to Fig. 3A of the main text where the interaction is repulsive. This could give additional information on the nature of the tip and on contrast formation in these measurements. The partial charge of the tip used here in Fig. S4B could be opposite to the one on the graphene nanoribbon's edge to explain the difference.

On the right hand side of Figure S4B two areas with dark edges have been marked with white arrows. These areas are defective and the graphene nanoribbon could be hydrogen-terminated at these areas similar to the defect shown in the main text.

\begin{figure}[h]
\centerline{
\includegraphics[width=16cm]{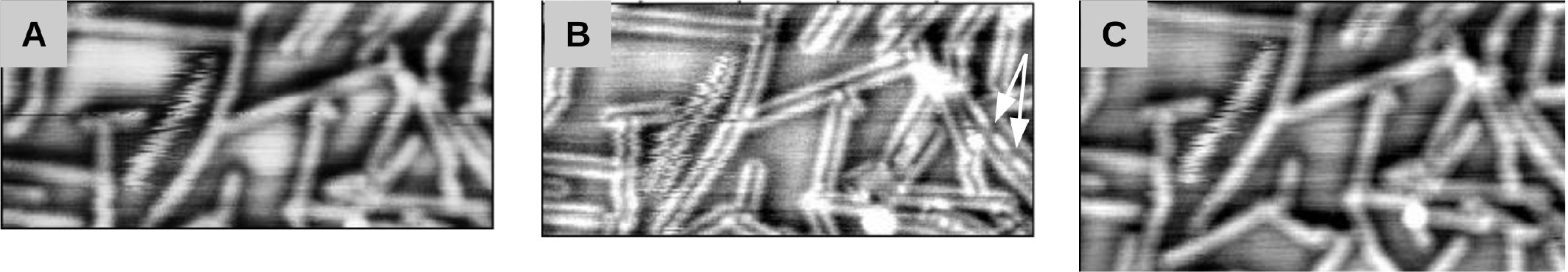}}
    \label{figS4}
\end{figure}

\paragraph{Fig. S4: Sequence of images showing a tip change and bright edge contrast.} These images were obained in standard frequency modulation scanning force microscopy mode with the bimodal oscillation switched off. They show the sample topography at constant frequency shift. ({\bf A}) Image obtained on graphene nanoribbons. ({\bf B}) After a tip change, the edges of the nanoribbons show bright contrast. ({\bf C}) The contrast changes back to its original state in the subsequent image.
$c_L = 37$ N/m, $f_0=331$ Hz, $\Delta f= -16$~Hz, $A = 6$ nm, $Q = 24 000$ and $T=115$ K.





\end{document}